\documentstyle[epsf,12pt]{article}
\def\vec#1{\mathchoice{\mbox{\boldmath$\displaystyle\bf#1$}}
{\mbox{\boldmath$\textstyle\bf#1$}}
{\mbox{\boldmath$\scriptstyle\bf#1$}}
{\mbox{\boldmath$\scriptscriptstyle\bf#1$}}}

\begin{document}
\begin{titlepage}
\begin{flushright}
PITHA 97/34 \\
September 1997
\end{flushright}
\vspace{0.8cm}
\begin{center}
{\bf\LARGE Signatures of Higgs Bosons \\}
{\bf\LARGE  in the Top Quark Decay Channel \\}
{\bf\LARGE  at Hadron Colliders\\}
\vspace{2cm}
{\bf W.\ Bernreuther, M.\  Flesch, and P.\ Haberl\footnote{
supported  by BMBF contract 057AC9EP.} }
\par\vspace{1cm}
Institut f.\ Theoretische Physik, RWTH Aachen, D-52056 Aachen, Germany
\par\vspace{3cm}
{\bf Abstract:}\\
\parbox[t]{\textwidth}
{ We analyze some signatures of neutral Higgs bosons, produced
in high energy proton (anti)proton collisions, which decay 
primarily to top quarks. This channel is clouded by a large 
irreducible $t\bar t$ background. We investigate the $t\bar t$ 
invariant-mass distribution in the lepton+jets decay mode of 
$t\bar t$ and a $t\bar t$ spin-spin correlation in the dilepton 
decay mode. At the LHC such Higgs bosons can be detected in the 
$t\bar t$ invariant-mass spectrum, and the spin-spin correlation 
studied by us is also sensitive to a Higgs boson if this 
particle has a sizeable pseudoscalar component. }
\end{center}
\vspace{2cm}
PACS number(s): 14.65.Ha, 14.80.Bn, 14.80.Cp
\end{titlepage}
%
%
\section{Introduction}
One of the most important physics issues at future hadron colliders,
in particular at the CERN Large Hadron Collider LHC, will be the search 
for Higgs bosons. While in the standard electroweak theory (SM) one 
neutral Higgs boson is associated with electroweak symmetry breaking, 
the situation is more complex in many extensions of the SM. For 
instance, the well-known  (non)supersymmetric two-Higgs doublet
extensions of the SM predict the existence of three neutral spin-zero 
particles: one CP-odd  and two CP-even Higgs bosons. The strategies 
to search for these particles \cite{rev} must take into account 
various possibilities. It may well be that the coupling to weak
vector bosons of one of these Higgs particles is suppressed -- or even
zero, as in the case of a pseudoscalar -- whereas its couplings to 
fermions is enhanced with respect to the SM Higgs couplings.
\par
In this article  we investigate the signatures of a  heavy Higgs boson
in the $t\bar t$  decay mode at hadron colliders. Because of the large 
nonresonant $t\bar t$ background this is known to be a very difficult
channel for the search of the SM Higgs boson. However, it is important to
study this channel in a more general theoretical setting because, as
mentioned above, Higgs particles with suppressed couplings to $W$ and 
$Z$ bosons but enhanced couplings to $t$ quarks may exist. Such a 
scenario was analyzed in \cite{Dicus}, and it was concluded in this 
paper that under favorable circumstances a 
scalar or pseudoscalar Higgs boson signal may be 
observable in the $t\bar t$ invariant-mass spectrum. Here we extend 
this analysis in several ways. We consider in the following a neutral 
Higgs boson $\varphi$ of arbitrary CP nature -- i.e., $\varphi$ can 
be a scalar, pseudoscalar, or a state of undefined CP parity. We 
investigate in sect.~3
the signal of $\varphi$  in the $t\bar t$ invariant-mass 
spectrum. We consider the case where this spectrum is reconstructed 
from the single lepton $t\bar t$ decay channel which is experimentally 
rather clean. Moreover, we exploit that in top quark production and 
decay the top spin is a ``good observable" in the sense that due to 
the extremely short lifetime of the $t$ quark the $t$ spin-polarization
and $t\bar t$ spin-spin correlations do not get severely diluted by
hadronization and can be traced by appropriate correlations among the 
final state particles into which $t$ and $\bar t$ decay [3--8].
The $t$ spin-polarization
and $t\bar t$ spin-spin correlations are sensitive to the Lorentz 
structure of the production vertex, and  the decay
$\varphi \to t\bar t$ leads to characteristic spin correlation signals
\cite{corr}. In sect.~4 we analyze for the double-lepton 
$t\bar t$ decay channel the effects of $\varphi$ production on 
$t\bar t$ spin-spin correlations, and we conclude in sect.~5.
\section{The $\varphi\to t\bar t$ channel including the QCD background}
In order to be completely general we consider a neutral Higgs boson 
$\varphi$ with unspecified CP parity. Its coupling to the top quark 
reads
\begin{equation}
{\cal L}_{Y}=-(\sqrt{2}G_F)^{1/2}m_t(a \bar{t}t+
\tilde{a} \bar{t}i\gamma_5 t)\varphi, 
\label{lagr}
\end{equation}
where $G_F$ is Fermi's constant, $m_t$ is the top mass, and 
$a, \tilde{a}$ are the reduced scalar and pseudoscalar couplings,
respectively. (For the SM Higgs boson $a=1$, $\tilde a=0$.)  If 
$a\tilde a\neq 0$ then $\varphi$ has undefined CP parity, signaling 
CP violation in the Higgs sector. This can occur, for instance,
in the two-Higgs doublet extension of the SM \cite{Wein}.
\par
In the following we consider a heavy $\varphi$ with mass 
$m_{\varphi}>2m_t$ with very small couplings to the weak vector bosons,
but unsuppressed couplings to $t$ quarks. Such a scenario may be 
realized in the supersymmetric extension of the SM \cite{Dicus} and
also in nonsupersymmetric two-doublet models \cite{rev}. Then, for  
$m_{\varphi}>2 m_t$  the mode $\varphi\to t\bar t$ determines the
total $\varphi$ decay width. To the extent that the decays of $\varphi$ 
to $W^+W^-$ and $ZZ$ are non-negligible the effects discussed in the 
next section become smaller.
\par
At high energy hadron colliders, for instance at the LHC, $\varphi$
will be produced by gluon-gluon fusion through a virtual top quark 
loop \cite{Georgi}. Scalar and pseudoscalar Higgs production by gluon 
fusion has been analyzed in detail in the literature \cite{QCD}.
The $\varphi\to t\bar t$ decay channel is affected by the large 
nonresonant $t\bar t$ background. The amplitudes 
$ g g\to \varphi \to t\bar t$ and $g g\to t\bar t$ interfere and 
produce at the parton level a characteristic peak-dip structure in the 
cross section \cite{Gaemers,Dicus,been} and in other observables 
\cite{BBra} if the $t \bar t$ invariant mass lies in the vicinity of 
the Higgs mass. As we are interested, apart from the $t\bar t$ 
invariant mass distribution, also in $t\bar t$ spin-spin correlation 
phenomena we have calculated the  $t\bar t$ production 
density matrices for the reactions $ q\bar q\to t\bar{t} $
and $ gg \to t\bar{t} $ by taking into account the QCD Born 
diagrams and the $s$ channel $\varphi$ production diagram.
The definitions and results are given in the appendix \cite{EW}.  
\par
The $t\bar t$ spin-spin correlations are sensitive to the dynamics of 
top quark production. This may be exemplified by the the following 
examples. It is known \cite{Hara,Arens,Brand} that close to threshold 
the gluon-gluon fusion amplitude at Born level yields a $t\bar t$ pair 
in a $^1\!S_0$ state. (The QCD Born amplitudes $q{\bar q}\to t\bar t$ 
leave the $t\bar t$ pair in a $^3\!S_1$ state.) Consider, on the other hand,
$g g\to\varphi\to t\bar t$. For a scalar and pseudoscalar $\varphi$ the
$t\bar t$ pair is in a $^3\!P_0$ and in a $^1\!S_0$ state, respectively.
Let us now consider the expectation value of the product 
${\bf s}_{t}\!\cdot\!{\bf s}_{\bar t}$ of the $t$ and $\bar t$ spins.
A simple computation yields that 
$\langle {\bf s}_{t}\!\cdot\!{\bf s}_{\bar t}\rangle=-3/4$
if $t\bar t$ is produced by gluon fusion at threshold, and
$\langle{\bf s}_{t}\!\cdot\!{\bf s}_{\bar t}\rangle = 1/4 (-3/4)$ 
if $t\bar t$ is produced by a (pseudo)scalar spin-zero boson 
and if the $g g\to t\bar t $ background  is ignored. 
In Fig.~1 we have plotted, as a function of the $t\bar t$ 
invariant mass, this expectation value for $g g\to t\bar t$ without 
and with the inclusion of the $s$ channel scalar,
respectively pseudoscalar Higgs boson exchange diagram. 
The Higgs boson mass $m_{\varphi}=354$ GeV was chosen to be close to 
the $t\bar t$ threshold of 350 GeV in order to exhibit the features 
discussed above. Above the threshold and resonance regions the Higgs
effect in $\langle{\bf s}_{t}\!\cdot\!{\bf s}_{\bar t}\rangle$ is 
actually more pronounced 
for a pseudoscalar than for a scalar, as can been seen from these 
figures. This will be further exemplified in sect.~4.
\par
The $t$ and $\bar t$ quarks 
auto-analyze their spins by their parity-violating weak decays.
We shall assume that $t\to W+b$ is the dominant decay mode, as 
predicted by the SM. It is well-known that the most efficient analyzer 
of the $t$ spin is the charged lepton from subsequent $W$ decay.
Its spin analyzer quality is more than twice as high as the $W$ or 
$b$ quark direction of flight. The corresponding
$t$ and $\bar t$ decay density matrices are given for instance in 
[23--25].
\par
In the following section we consider two types of $t\bar t$ decay 
channels: first the ``single lepton channel", where the 
$t$ quark decays semileptonically and the $\bar t$ nonleptonically,
\begin{equation}
t+\bar t \to l^+\nu_l b + q \bar{q}' \bar b \;, 
\end{equation}
and vice versa. These channels have a good signature for top 
quark identification. Moreover, the $t$ and $\bar t$ momenta can be 
reconstructed up to an ambiguity which can be resolved 
statistically \cite{Lad}. These processes are therefore suitable for
determining the $t \bar t$ invariant-mass spectrum.
Second we consider the processes -- which we shall call below 
``double lepton channel" -- where both $t$ and 
$\bar t$ decay semileptonically,
\begin{equation}
t+\bar t \to l^+\nu_l b +  l'^-\bar{\nu}_{l'}\bar b \;. 
\end{equation}
As mentioned above, correlations among 
the directions of flight of the charged leptons are best suited to 
analyze $t\bar t$ spin correlations. We will therefore study
angular correlations in the double lepton channel.
\par
For obvious reasons we demand the lepton in a semileptonic
top decay to be either an electron or a muon.
The SM predicts, to good approximation,
a fraction of 24/81 of all $t\bar t$ events to decay into the 
single lepton channel and 4/81 into the double lepton channel.
\section{The $t\bar t$ invariant mass spectrum}
Throughout our calculation we use the narrow width 
approximation for $t\bar t$ production and decay, which is 
justified because of $\Gamma_t/m_t,\;\Gamma_W/m_W \ll 1$ 
and because we are primarily interested in 
normalized distributions and 
expectation values of observables. Moreover we take
the leptons and the light quarks including the $b$ quark to be
massless. Let us start with writing down our ``master formula"
for the cross section measure for the double lepton channel:
\begin{eqnarray}
\label{dsigma}
\int d\sigma \!\!\!\!\!&(&\!\!\!\!\!\!pp \to t\bar t X
\to l^++\nu_l+b +l'^-+\bar{\nu}_{l'}+{\bar b}+X) \nonumber \\
&=& {\cal N} \sum_{\lambda=q,\bar q,g} 
\int_0^1 \!dx_1 \int_0^1 \!dx_2 \; N_\lambda(x_1)
  N_{\bar \lambda}(x_2) \Theta(\hat{s}-4m_t^2) \nonumber \\
&& \times \;
   \frac{\alpha_s^2\beta}{\hat s}\int d\Omega_{\hat{\vec k}}\;
   \frac{1}{\eta}\int_\mu^1 dy_+ y_+(1-y_+)
   \frac{1}{\eta}\int_\mu^1 dy_- y_-(1-y_-) \nonumber \\
&& \times 
   \int\frac{d\Omega_{\hat{\vec q}_+}}{4\pi}
   \int\frac{d\Omega_{\hat{\vec q}_-}}{4\pi}
   \Bigg\{A^{(\lambda)}
   + B^{(\lambda)}\left( \hat{\vec k}\!\cdot\!\hat{\vec q}_+
      +\hat{\vec k}\!\cdot\!\hat{\vec q}_- \right) \nonumber \\
&& -\bigg[ c_0^{(\lambda)}\hat{\vec q}_+\!\!\cdot\!\hat{\vec q}_-
   + c_2^{(\lambda)}\hat{\vec k}\!\cdot\!
   (\hat{\vec q}_+\!\!\times\!\hat{\vec q}_-) 
   + c_4^{(\lambda)}(\hat{\vec p}\!\cdot\!\hat{\vec q}_+)
       (\hat{\vec p}\!\cdot\!\hat{\vec q}_-)    \\
&& + c_5^{(\lambda)}(\hat{\vec k}\!\cdot\!\hat{\vec q}_+)
                    (\hat{\vec k}\!\cdot\!\hat{\vec q}_-)
   + c_6^{(\lambda)}\left((\hat{\vec p}\!\cdot\!\hat{\vec q}_+)
                          (\hat{\vec k}\!\cdot\!\hat{\vec q}_-)+
                          (\hat{\vec p}\!\cdot\!\hat{\vec q}_-)
                          (\hat{\vec k}\!\cdot\!\hat{\vec q}_+)
   \right)\bigg]\Bigg\} \nonumber \\
&& \times \int_0^\infty dE_b\;
   \delta\left(E_b-\frac{m_t^2\!-\!m_W^2}{2m_t}\right)
   \int\frac{d\Omega_{\hat{\vec q}_b}}{2\pi}\;
   \delta\left(\hat{\vec q}_+ \!\!\!\cdot\!\hat{\vec q}_b -
   \frac{2\mu -y_+(1+\mu)}{y_+(1-\mu)}\right) \nonumber \\
&& \times \int_0^\infty dE_{\bar b}\;\delta\left(E_{\bar b}-
   \frac{m_t^2\!-\!m_W^2}{2m_t}\right)
   \int\frac{d\Omega_{\hat{\vec q}_{\bar b}}}{2\pi}\;
   \delta\left(\hat{\vec q}_- \!\!\!\cdot\!\hat{\vec q}_{\bar b} -
   \frac{2\mu -y_-(1+\mu)}{y_-(1-\mu)}\right) \;. \nonumber 
\end{eqnarray}
In the first line on the r.h.s.\ ${\cal N}={\cal B}(t\to bl\nu_l)^2$
is the square of the semileptonic $t$ branching ratio,
and $N_\lambda(x_1)$, $N_{\bar \lambda}(x_2)$ are the
parton distribution functions.
The next four lines in eq.\ (\ref{dsigma})
represent the differential cross section at the parton level,
written in terms of the $t$ momentum
direction $\hat{\vec k}$ in the parton c.m.\ system,  
the lepton momentum directions $\hat{\vec q}_\pm$,
and the normalized lepton energies $y_\pm=2E_\pm/m_t$,
defined in the $t$ or $\bar t$ rest systems,
respectively. The minimal value of the normalized lepton 
energies is $\mu=m_W^2/m_t^2$.
Forefactors are the strong coupling constant $\alpha_s$
and the kinematic factor $\beta=(1-4m_t^2/\hat s)^{1/2}$ with
the parton c.m.\ energy $\hat{s}=x_1x_2s$.
The factor $\eta=(1-\mu)^2(1+2\mu)/6$ is chosen such that 
the lepton energy integrations are normalized to unity.
\par
The matrix element contains the coefficients
$A^{(\lambda)}$, $B^{(\lambda)}$, 
$c_0^{(\lambda)}\ldots c_6^{(\lambda)}$, 
which depend only on $\hat{s}$ 
and the cosine of the angle between $\hat{\vec k}$ and the beam
direction, 
$z=\hat{\vec p}\!\cdot\! \hat{\vec k}$. They are listed in
the appendix. Note that only $A^{(\lambda)}$
contributes to the rate, while the other coefficients
lead to $t$ and $\bar t$ spin polarization and spin-spin
correlations. In particular, the total cross section
for the parton subprocess $\lambda\bar{\lambda}\to t\bar t$ is
simply
\begin{equation}
\hat{\sigma}^{(\lambda)}(\hat s) = \frac{\alpha_s^2\beta}{\hat s}
\int d\Omega_{\hat{\vec k}} A^{(\lambda)}
(\hat{s},z)\;,
\end{equation}
which can be computed analytically.
\par
The last two lines in eq.\ (\ref{dsigma})
finally give the distribution of 
$b$ ($\bar b$) energies $E_{b}$ ($E_{\bar b}$) and
directions $\hat{\vec q}_b$  ($\hat{\vec q}_{\bar b}$),
defined again in the $t$ ($\bar t$) rest system.
The four $\delta$--functions have their origin
in the top and $W$ propagators, for which we have used the
narrow width approximation.
One can see that once the lepton momenta are given, the 
$b$ ($\bar b$) momenta are fixed up to an azimuthal angle
around the accompanying lepton, on which the matrix element does
not depend.
\par
The cross section for $p\bar p$ collisions is simply obtained 
from eq.\ (\ref{dsigma}) by substituting the parton distribution 
function $N_{\bar \lambda}(x_2)$ with $\bar{N}_{\bar\lambda}(x_2)$,
which enhances the incoherent quark annihilation background.
Note that eq.\ (\ref{dsigma}) can be as well applied to reactions
where one or both of the $t$ quarks decay hadronically;
in this case the lepton and neutrino momenta are to be interpreted
as the momenta of the (light) quarks into which the $W$ decays,
and one has to supplement the appropriate branching ratios.
\par 
Let $M_{t\bar t}=\sqrt{(k_t+k_{\bar t})^2}$ be the
invariant mass of the $t\bar t$ system. The spectrum
$d\sigma/dM_{t\bar t}$ is then obtained \cite{Dicus} from
multiplying the total parton cross sections 
$\hat{\sigma}^{(\lambda)}$ at $\hat{s}=M^2_{t\bar t}$ with the
so-called luminosity functions
\begin{equation} 
L^{(\lambda)}(M_{t\bar t},s)=\frac{2\tau}{\sqrt{s}}
\int_\tau^{1/\tau}\frac{d\zeta}{\zeta}
N_{\lambda}(\tau\zeta)
N_{\bar \lambda}(\frac{\tau}{\zeta})
\end{equation}
where $\tau=M_{t\bar t}/\sqrt{s}$. However, 
this approach does not account for the experimental
difficulty of reconstructing the top momenta, which is
present in the single lepton channel, and it is not a 
priori clear to what extent a misidentification might
disturb the signal. Apart from that,
the implementation of cuts is not straightforward.
\par
We have therefore taken eq.\ (\ref{dsigma}) as the starting point 
for writing a Monte Carlo generator. Importance sampling is
performed in the variables $x_1$ and $x_2$, where we expect
the biggest variations (due to the threshold behavior of the
parton cross section and the form of the parton distribution
functions). As can be seen from eq.\ (\ref{dsigma}),
the lepton energies can be simulated independently.
Moreover, the construction of the $b$ and $\bar b$ momenta
requires only two (equally distributed) azimuthal angles,
as mentioned above.
\par
The obstacle to reconstructing both the $t$ and $\bar t$ momenta 
in the single lepton channel is the missing neutrino momentum.
Four-momentum conservation and the various on-shell conditions
determine the latter as the solution of a quadratic
equation up to a twofold ambiguity. In ref.\ \cite{Lad}
a strategy was described how to improve on the odds for 
picking the correct solution: In a first step, cuts are applied
on the rapidity of the top quark and the $W$ boson on the
hadronic side. Next one tests whether the reconstructed 
$x_1,x_2$ lie in the interval $[0,1]$. Finally one chooses 
that neutrino momentum which -- when combined with the lepton
momentum -- gives a value for the invariant $l$-$\nu_l$ mass
which is closest to the physical $W$ mass.
This procedure can raise the probability for a correct 
identification to 86--89\%.
\par
It is a simple matter to implement this algorithm into our 
Monte Carlo generator: Although we know the 
correct neutrino momentum, we compute for each event the
wrong solution and construct a ``wrong" top momentum,
which we select in 11--14\% of all cases for a determination
of $M_{t\bar t}$. In addition we apply the following cuts:
\begin{equation}
|y(t)|\le 3\;,\quad\quad p_T(l)\ge 20\,{\rm GeV}\;,
\label{cuts}
\end{equation}
for the rapidity of the top quark and the transverse
momenta of the leptons. We will keep these cuts
throughout below, also for the double lepton channel
in the next section. For the numerics we put $m_t=175$ GeV
and $m_W=80.4$ GeV, and we set the partial widths
$\Gamma(\varphi\to W^+W^-,ZZ)$ to zero. We took the parton 
distribution functions (PDF) from ref.\ \cite{GRV}, evaluated 
at the factorization scale
$\Lambda=m_t$, and convinced ourselves that our results do not 
change significantly if we vary $\Lambda$ between $m_t/2$
and $2m_t$ or work with other PDF sets \cite{MRSCTEQ}.
\par
Fig.~2 shows the resulting normalized invariant $t\bar t$ 
mass spectrum $\sigma^{-1}d\sigma/dM_{t\bar t}$
for $p\bar p$ collisions at $\sqrt{s}=4$ TeV
in bins of 10 GeV, starting at the threshold of
$2m_t=350$ GeV. In grey we show the result from the
QCD background and superimpose as solid lines the curves
which we obtained including the production of a Higgs boson 
of $m_\varphi=400$ GeV. The respective values of the 
$\varphi t\bar t$ couplings are chosen to be:
2a) $a=1$, $\tilde a=0$ (scalar),
2b) $a=0$, $\tilde a=1$ (pseudoscalar), and
2c) $a=\tilde a=1$ (undefined CP parity).
Each of the plots in Fig.~2 was produced with 50 000 events.
This number  results from assuming $p\bar p$ collisions
with 10 ${\rm fb}^{-1}$ integrated luminosity and a 
detection efficiency of 1/3 for the single lepton channel.
\par
For the Yukawa couplings used in Fig.~2 the $\varphi$
resonance becomes broader if $\tilde a \neq 0$. 
As a consequence,
the contribution to $gg\to t\bar t$ of a $\varphi$ boson with
pseudoscalar component is, markedly above 
$\hat{s}>m^2_\varphi$, more visible than that of a 
pure scalar. The effect in the normalized spectra shown in Fig.~2
is therefore more pronounced for a $\varphi$ boson with
$\tilde a \neq 0$ than for a pure scalar $\varphi$.
One can see that the typical peak-dip structure 
in the $M_{t\bar t}$ spectrum described in \cite{Dicus} 
survives at the level of the $t$ and $\bar t$ decay products, 
even if one implements 
realistic cuts and misidentifies the top quark momentum 
in $\sim11$\% of all events.
\par
For the upgraded Tevatron an integrated luminosity of 
10 ${\rm fb}^{-1}$ is eventually projected, but at
$\sqrt{s}=2$ TeV. At this c.m.\ energy no significant
signal of a $\varphi$ boson with the Yukawa couplings
of Fig.~2 remains above the $t\bar t$ background.
\par
The situation clearly improves if we go to $pp$ collisions at
LHC energies, one reason being the suppression of the quark
annihilation background. In Fig.~3 we show the same series of
plots as in Fig.~2, but now for $pp$ at 14 TeV. These plots
were produced with $1.2\!\times\!10^6$ events. This is a 
conservative number for the LHC with integrated luminosity
of 100 ${\rm fb}^{-1}$. Qualitatively the 
signals are comparable to those in Fig.~2, but cleaner due to 
the larger number of events. The clear signals are of course due
to the fact that we chose the Higgs mass $m_\varphi=400$ GeV to
be close to the threshold of $2m_t=350$ GeV. To study the effect
of different Higgs boson masses, we show in Fig.~4 a series
of invariant mass distributions obtained for $m_\varphi=375$, 
400, 450 and 500 GeV. In these plots, $\varphi$ was taken to 
be pseudoscalar with Yukawa couplings $a=0$ and $\tilde a=1$.
\par
Let us briefly discuss the statistical significance of our
results, for the case of a pseudoscalar with mass
500 GeV (cf.\ Fig.~4d). Only 84.2\% of the $1.2\!\times\! 10^6$ 
events pass the cuts (\ref{cuts}) 
and are filled into the histogram.
If we consider the contents of the five bins below $m_\varphi$,
i.e.\ the $M_{t\bar t}$ interval between 450 GeV and 500 GeV, we 
find $n_0=$161 765 events for the background and  164 392 events 
if the $\varphi$ exchange is included, which is a 1.6\% effect.
Due to the large number of events this is a significant increase: 
With a 1 s.d.\ uncertainty of about $\sqrt{n_0}\simeq402$, 
this translates into a 6.5 $\sigma$ effect.
The statistical significance is, of course, higher for the
curves depicted in Figs.~4a,b,c. 
\par
The Higgs effect decreases
if the decays $\varphi\to W^+W^-,\,ZZ$ are non-negligible.
If the couplings of $\varphi$ to the weak vector bosons
are of SM strength then, using the same number of events,
$\varphi$ masses, and Yukawa couplings as in Figs.~4a--d, 
the $\varphi$ resonance effect in the $M_{t\bar t}$ spectrum
becomes insignificant -- except possibly for a $\varphi$ with
pseudoscalar component and a mass close to $2m_t$.
\section{Spin-spin correlations of the $t\bar t$ system}
We have also studied the impact of the $s$-channel 
$\varphi$-exchange diagram on various correlations
among the final state particles.
We have concentrated on the double leptonic decay mode and
investigated several observables involving the lepton
energies and momenta.
In particular, the spin-spin correlation
$\langle{\bf s}_{t}\!\cdot\!{\bf s}_{\bar t}\rangle$
discussed in sect.~2 suggests to investigate 
the observable
\begin{equation}
{\cal O} = \hat{\vec{Q}}_+\!\!\cdot\!\hat{\vec{Q}}_- \;,
\end{equation}
where $\hat{\vec{Q}}_\pm$ are the unit vectors of the
lepton momenta measured in the hadron c.m.\ system.
In other words, ${\cal O}$ is the cosine of the angle
between the directions of flight of the two leptons
as seen in the laboratory frame. Recall that the lepton momenta 
${\vec q}_\pm$ in $d\sigma$ (\ref{dsigma}) are defined in the 
$t$ and $\bar t$ rest systems and thus related 
to $\vec{Q}_\pm$ by two (rotation-free) boosts.
\par
The grey area in Fig.~5 shows the distribution of ${\cal O}$ 
for $pp$ collisions at 6 TeV, as obtained from the background 
only. One can see that event topologies where the two leptons 
are back to back or parallel are slightly preferred.
In the same figure we show as solid line the resulting 
distribution if we assume the presence of a pseudoscalar
Higgs boson with mass $m_\varphi=400$ GeV and Yukawa 
couplings $a=0$, $\tilde a=2$. In this case the distribution
is shifted towards positive values of ${\cal O}$.
\par
In order to quantify this effect one could study the
asymmetry between the event numbers with positive and 
negative ${\cal O}$. The same information is of course 
contained in the expectation value $\langle{\cal O}\rangle$.
The sensitivity of $\langle{\cal O}\rangle$ to the couplings 
$a$ and $\tilde a$ is determined by the 1 s.d.\ width
$\delta {\cal O}=\sqrt{(\langle {\cal O}^2 \rangle
-\langle {\cal O} \rangle^2 )/N}$
where $N$ is the number of events in the sample. 
\par
In Fig.~6 we show the theoretical expectation for 
$\langle{\cal O}\rangle$, for $pp$ collisions from 4 to 14 TeV.
The solid line and the dark grey band represent for a given
c.m.\ energy $\sqrt{s}$ the center value and the 3 $\sigma$ interval
$[\langle {\cal O}\rangle -3\delta {\cal O},
\langle {\cal O}\rangle +3\delta {\cal O}]$.
The width $\delta{\cal O}$ was computed with $N_{ll}=2\!\times\!10^5$ 
events, which is 1/6 of the single lepton 
events used in the previous section. 
Also shown in Fig.~6 -- as dashed line with light grey band --
is the effect on $\langle{\cal O}\rangle$ and $\delta{\cal O}$
caused by a $m_\varphi=400$ GeV Higgs boson with 
pseudoscalar couplings $a=0$, $\tilde a=2$. One can see that 
this leads to an increase of $\langle{\cal O}\rangle$
with a significance above the 3 $\sigma$ level. With the same 
coupling strength and mass, a scalar Higgs boson (i.e.\ $a=2$, 
$\tilde a=0$) does not yield a significant signal.
This is expected from the discussion in sect.~2.
\par
Both Figs.~5 and 6 were produced with the cuts eq.\ (\ref{cuts}).
We observed that the signal gets enhanced by
the cut on the lepton transverse momentum.
\par
One might think that the correlation effects increase if one works 
with the lepton directions $\hat{\vec q}_\pm$ measured in the 
$t$ and $\bar t$ rest systems. Although we consider the 
double lepton channel, it is possible to reconstruct
both the $t$ and $\bar t$ momenta \cite{Chang}, albeit
with a certain loss in efficiency. We have therefore 
investigated also the observable
\begin{equation}
{\cal O}' = \hat{\vec{q}}_+\!\!\cdot\!\hat{\vec{q}}_- \;.
\end{equation}
The results for $\langle{\cal O}\rangle$ and 
$\langle{\cal O}'\rangle$ are qualitatively similar. 
However, in the case of $\langle{\cal O}'\rangle$ already
the 2 $\sigma$ bands overlap, which means that the observable
${\cal O}'$ is less sensitive to the Yukawa couplings
$a$ and $\tilde a$ than ${\cal O}$.
\par
Finally we mention that neutral Higgs 
sector CP violation, i.e.\ $a\tilde a\neq 0$, can
be tested with appropriate asymmetries
and correlations \cite{BBra,Peskin,Haberl,BM,Grz}.
\section{Conclusions}
Extensions of the SM predict a number of neutral Higgs bosons 
$\varphi$. It is possible that one -- or several -- of these bosons 
have very small couplings to $W$ and $Z$ bosons but unsuppressed 
couplings to top quarks. In this paper we have considered the 
resonant production of such particles $\varphi$ in the reaction  
$p + p ({\bar p}) \to t +{\bar t} + X$. We have studied the effect 
of $\varphi$ on the $t\bar t$ invariant-mass distribution
obtained through the single lepton channel. At the upgraded Tevatron
($\sqrt s$ = 2 TeV) with integrated luminosity of 10 fb$^{-1}$ the 
signal of $\varphi$ is insignificant -- assuming that the Yukawa 
couplings of $\varphi$ do not exceed SM strength -- because of 
limited statistics and the  large $t\bar t$ background from
quark-antiquark annihilation. For $p\bar p$ collisions at 
$\sqrt s$ = 4 TeV, assuming again 10 fb$^{-1}$ of integrated 
luminosity, $\varphi$ bosons become visible in the $t\bar t$ 
invariant-mass spectrum for masses $m_{\varphi}$ around 2 $m_t$.
For $p p$ collisions at LHC energies and luminosities 
our analysis suggests that $\varphi$ production is detectable
for a range of masses $m_{\varphi}$.
\par
If such bosons should be found the analysis of their properties can 
be supplemented by analyzing $t$ and $\bar t$ spin-polarization and 
spin-spin correlation phenomena in the $\varphi\to t\bar t$ decay 
channel. The spin-spin correlation that we have studied is, at LHC 
energies, sensitive to a resonant $\varphi$ contribution if $\varphi$ 
has a sizeable pseudoscalar coupling to top quarks. This and other 
correlations and asymmetries should be useful in eventually pinning 
down the nature of these bosons. Our study was based on the lowest
order parton matrix elements. Detailed analyses eventually require
the inclusion of QCD radiative corrections, of hadronization
and detector effects.
\subsubsection*{Acknowledgments}
We thank A.~Brandenburg for discussions.
\newpage
\section*{Appendix}
\renewcommand{\theequation}{A.\arabic{equation}}
\setcounter{equation}{0}
We define a decay density matrix $\rho$ for the process
$t\rightarrow \ell^+\nu_\ell \;b$ in the narrow width 
approximation for the $W$ boson by
\begin{equation}
\frac{\pi}{m_W\Gamma_W}\delta(p_W^2-m_W^2)\rho_{\alpha'\alpha}=\!\!\!
\sum_{\ell^+\nu_\ell b \;{\rm\scriptscriptstyle spins} }\!\!\!
\langle t_{\alpha'}|{\cal T}^\dagger|\ell^+\nu_\ell b \rangle\;
\langle \ell^+\nu_\ell b|{\cal T}|t_{\alpha}\rangle\;,
\end{equation}
where $p_W$, $m_W$ and $\Gamma_W$ are the four-momentum, mass and 
width of the intermediate $W^+$,
and $\alpha$, $\alpha'$ are the $t$ spin indices.
The explicit form of $\rho$, conveniently evaluated in the 
top quark rest system, can be found e.g.\ in \cite{Haberl}.
The matrix $\bar\rho$ for the conjugate decay 
$\bar t\rightarrow \ell^-\bar{\nu}_\ell \;\bar b$
is obtained from $\rho$ by simple replacements.
\par
Similarly, we define the production density matrices for the 
partonic process
$\lambda\bar\lambda\rightarrow t\bar t$ ($\lambda=q,g$) as
\begin{equation}
R^{(\lambda)}_{\alpha\alpha' ,\beta\beta'}=
\frac{1}{g_s^4}\frac{1}{n_{(\lambda)}}
\sum_{{{\rm\scriptscriptstyle colors} \atop 
{\rm\scriptscriptstyle initial}\;
{\rm\scriptscriptstyle spins} }}
\langle t_\alpha\bar t_\beta |{\cal T}|
\,\lambda\bar\lambda\,\rangle\;
\langle \,\lambda\bar\lambda\,|{\cal T}^\dagger|
t_{\alpha'}\bar t_{\beta'}\rangle\;,
\end{equation}
where the factor $n_{(\lambda)}$ averages over spin and color
of the initial state partons: $n_{(q)}=(2N_C)^2=36$,
$n_{(g)}=(2(N^2_C-1))^2=256$.
The factor $g_s^4$ is taken out for convenience.
A general discussion of the symmetry properties of these matrices
and their decomposition in the $t$ and $\bar t$ spin spaces
is given in \cite{BBra}. Here we consider only QCD Born amplitudes
and the lowest order $gg\to\varphi\to t\bar t$ amplitude.
Therefore several of the coefficients in the general
decomposition vanish, and the matrices $R^{(\lambda)}$ are
of the form
\par
\hfill\parbox{13.4cm}{
\begin{eqnarray*}
R^{(\lambda)}_{\alpha\alpha',\beta\beta'}&=&
A^{(\lambda)}\delta_{\alpha\alpha'}\delta_{\beta\beta'}
+B^{(\lambda)} \hat{k}_i\left((\sigma^i)_{\alpha\alpha'}
\delta_{\beta\beta'}-\delta_{\alpha\alpha'}
(\sigma^i)_{\beta\beta'} \right)\\
&&+C^{(\lambda)}_{ij}(\sigma^i)_{\alpha\alpha'}
(\sigma^j)_{\beta\beta'}  \;,
\end{eqnarray*} }
\hfill\parbox{0.8cm}{\begin{eqnarray}  \end{eqnarray} }
\par\noindent
where $\sigma^i$ are the Pauli matrices. In our case, the 
tensors $C^{(\lambda)}_{ij}$ have the structure
\begin{equation}
C^{(\lambda)}_{ij}=c^{(\lambda)}_0\,\delta_{ij}
+c_2^{(\lambda)}\,\varepsilon_{ijl}\,\hat{k}_{l}
+c_4^{(\lambda)}\,\hat{p}_{i}\hat{p}_{j}
+c^{(\lambda)}_5\,\hat{k}_{i}\hat{k}_{j}
+c^{(\lambda)}_6\left(\hat{p}_{i}\hat{k}_{j}
   +\hat{p}_{j}\hat{k}_{i}\right) \;.
\label{Cij}
\end{equation} 
Here $\hat{\vec k}$ and $\hat{\vec p}$ are the unit vectors
of the momenta of the top quark and of the initial parton 
$\lambda$, respectively, defined in the parton c.m.\ system. 
(Neglecting transverse parton momenta implies that
$\hat{\vec p}$ is equal to the
direction of the proton beam in the laboratory frame.)
The squared matrix element of the reaction 
$\lambda\bar{\lambda}\to t\bar t\to l^+\nu_lb+l'^-\bar{\nu}_{l'}\bar b$
is then proportional to
\begin{equation}
{\rm Tr}\;[\rho R^{(\lambda)}\bar{\rho}]\equiv\rho_{\alpha'\alpha}
R^{(\lambda)}_{\alpha\alpha',\beta\beta'}\bar{\rho}_{\beta'\beta}\;,
\end{equation}
which leads -- after integration over the phase space and
folding with parton distribution functions -- to the
representation of the differential cross section $d\sigma$
given in eq.\ (\ref{dsigma}). 
\par
The matrix $R^{(q)}$ gets only contributions from the
QCD background, leading to the coefficients
\par\noindent
\parbox{6.7cm}{
\begin{eqnarray*}
A^{(q)}   &=& N_q \Big[ 1 - \frac{\beta^2}{2}(1-z^2) \Big]\;, \\
c_0^{(q)} &=& N_q\Big[\hphantom{1}-\frac{\beta^2}{2}(1-z^2)\Big]\;, \\
c_4^{(q)} &=& N_q \;,
\end{eqnarray*} }
\hfill\parbox{6.7cm}{
\begin{eqnarray*}
c_5^{(q)} &=& N_q \left[ \beta^2 + z^2(1-x)^2 \right] \;,\\[2mm]
c_6^{(q)} &=& N_q \left[ - z(1-x) \right]\;, \\[2mm]
B^{(q)} &=& c_2^{(q)} \;\,=\;\, 0\;,
\end{eqnarray*} }
\hfill\parbox{0.8cm}{\begin{eqnarray} \label{qcoeff} \end{eqnarray} }
\par\noindent
with
\begin{equation}
x=\frac{2m_t}{\sqrt{\hat s}}\;,\quad\quad
\beta=\sqrt{1-x^2}\;,\quad\quad
z=\hat{\vec p}\cdot\hat{\vec k} \;,
\end{equation}
and the normalization factor
\begin{equation}
N_q = \frac{1}{4}\frac{C_F}{C_A}\;. \\
\end{equation}
As usual, the $SU(N_C)$ color factors are  
$C_F=(N_C^2-1)/(2N_C)$, $C_A=N_C$.
\par
We split the coefficients of the gluon fusion density matrix 
$R^{(g)}$ into contributions from the QCD background and
contributions involving the $\varphi$ exchange diagram:
\begin{equation}
A^{(g)}=A^{(g)}_{\rm Born}+A^{(g)}_{\varphi}\;,
\end{equation}
and similar for the other coefficients. For the background
part we obtain
\par
\hfill\parbox{13.4cm}{
\begin{eqnarray*}
A^{(g)}_{\rm Born} &=& 
  N_g^{(0)} \left[ 1+x^2-\frac{1-\beta^2z^2}{2}
  - \frac{x^4}{1-\beta^2z^2} \right] \;,\\
c^{(g)}_{0,{\rm Born}} &=& 
  N_g^{(0)} \left[ \hphantom{1+}\;\, x^2 - \frac{1-\beta^2z^2}{2}
  - \frac{x^4}{1-\beta^2z^2} \right] \;,\\
c^{(g)}_{4,{\rm Born}} &=& 
  N_g^{(0)} \left[ \frac{\beta^2(1-z^2)}{1-\beta^2z^2} \right] \;,\\
c^{(g)}_{5,{\rm Born}} &=& 
  N_g^{(0)} \frac{1-x}{1+x} \left[ 1 + 2x(1 + x) + \beta^2z^2 
    -\frac{x^2(3+4x+2x^2)}{1-\beta^2z^2}  \right] \;,\\
c^{(g)}_{6,{\rm Born}} &=& 
  N_g^{(0)}\left[-z(1-x)\right]
  \left[\frac{\beta^2(1-z^2)}{1-\beta^2z^2}\right] \;, \\
B^{(g)}_{{\rm Born}} &=& c^{(g)}_{2,{\rm Born}} \;\,=\,\; 0\;.
\end{eqnarray*} }
\hfill\parbox{0.8cm}{\begin{eqnarray}  \end{eqnarray} }
\par\noindent
For the contributions from $\varphi$ exchange it is 
convenient to define the following two functions, 
\par
\hfill\parbox{13.4cm}{
\begin{eqnarray*}
D &=& (-2+\beta^2\hat{s}C_0)\frac{\hat s}{\hat s-m_\varphi^2
+im_\varphi\Gamma_\varphi}  \;, \\ 
\tilde{D} &=& \hat sC_0\frac{\hat s}{\hat s-m_\varphi^2
+im_\varphi\Gamma_\varphi}\;,
\end{eqnarray*} }
\hfill\parbox{0.8cm}{\begin{eqnarray}  \end{eqnarray} }
\par\noindent
which contain the loop integral
\begin{equation}
C_0\equiv C_0(p_1,p_2,m_t,m_t,m_t)=\frac{1}{2\hat s}\left[
\ln\left(\frac{1+\beta}{1-\beta}\right)-i\pi\right]^2
\end{equation}
as well as the Breit--Wigner form of the $\varphi$ propagator.
With the help of $D$ and $\tilde D$ the
remaining contributions can be written in a compact way:
\par
\hfill\parbox{13.4cm}{
\begin{eqnarray*}
A^{(g)}_{\varphi} &=&  N_g^{(2)} \left[
    \beta^2a^2 {\rm Re}(D) +{\tilde a}^2 {\rm Re}(\tilde D)\right] \\
   & +& N_g^{(4)}  (\beta^2a^2+{\tilde a}^2)\left(a^2|D|^2 
    + {\tilde a}^2|\tilde D|^2 \right) \;, \\
B^{(g)}_{\varphi} &=& N_g^{(2)} (\beta a{\tilde a})
   {\rm Im}(D-\tilde D)  \;,\\
c^{(g)}_{0,\varphi} &=&  N_g^{(2)} \left[   
    \beta^2a^2 {\rm Re}(D) - {\tilde a}^2 {\rm Re}(\tilde D) \right] \\
  &+& N_g^{(4)}  (\beta^2a^2-{\tilde a}^2)\left(a^2|D|^2
    + {\tilde a}^2|\tilde D|^2 \right)  \;,\\
c^{(g)}_{2,\varphi} &=& N_g^{(2)}(-\beta a{\tilde a}){\rm Re}(D+\tilde D)  
  + N_g^{(4)} (-2\beta a{\tilde a}) \left( a^2|D|^2 
    + {\tilde a}^2|\tilde D|^2 \right) \;,\\
c^{(g)}_{5,\varphi} &=&
   N_g^{(2)}  (-2\beta^2a^2)  {\rm Re}(D) 
 + N_g^{(4)} (-2\beta^2a^2) \left( a^2|D|^2 
   + {\tilde a}^2|\tilde D|^2 \right) \;, \\
c^{(g)}_{4,\varphi} &=& c^{(g)}_{6,\varphi} \;\,=\,\; 0\;.
\end{eqnarray*} }
\hfill\parbox{0.8cm}{\begin{eqnarray}  \end{eqnarray} }
\par\noindent
Here $a$ and $\tilde a$ are the reduced scalar and pseudoscalar
Yukawa couplings defined in eq.\ (\ref{lagr}). The normalization
factors are 
\par
\hfill\parbox{13.4cm}{
\begin{eqnarray*}
N_g^{(0)} &=& \frac{1}{2} \left( \frac{1}{C_A} 
    \frac{1}{1-\beta^2z^2} -\frac{1}{4C_F} \right) \;,\\
N_g^{(2)} &=& \frac{1}{2} \frac{1}{C_FC_A} \frac{1}{32\pi^2} 
     \left(\frac{m_t}{v}\right)^2  \frac{x^2}{1-\beta^2z^2} \;,\\
N_g^{(4)} &=& \frac{1}{2} \frac{1}{C_F} \left( \frac{1}{32\pi^2}
    \right)^2  \left(\frac{m_t}{v}\right)^4 x^2  \;,
\end{eqnarray*} }
\hfill\parbox{0.8cm}{\begin{eqnarray} \label{gnorm} \end{eqnarray} }
\par\noindent
with $v=(\sqrt{2}G_F)^{-1/2}=246$ GeV. 
The coefficients given in this appendix agree with those obtained 
in \cite{BBra}. The cross sections for the partonic reactions
are determined by $A^{(\lambda)}$, and our expression for
$d\hat{\sigma}(gg\to t\bar t)/dz$ agrees with \cite{Dicus}.
\newpage
\newpage
\centerline{\bf Figure 1}
\begin{figure}[h]
\unitlength1.0cm
\begin{picture}(15.,15.)
\put(-1.,10.9){\makebox(2.,1.){\large 
$\langle\, {\bf s}_{t}\!\cdot\!{\bf s}_{\bar t} \,\rangle$}}
\put(7.2,0.6){\makebox(2.,1.){\large $\sqrt{\hat s}\;\;$ [GeV]}}
\put(0.3,1.0){
\epsfysize=15.cm
\epsffile{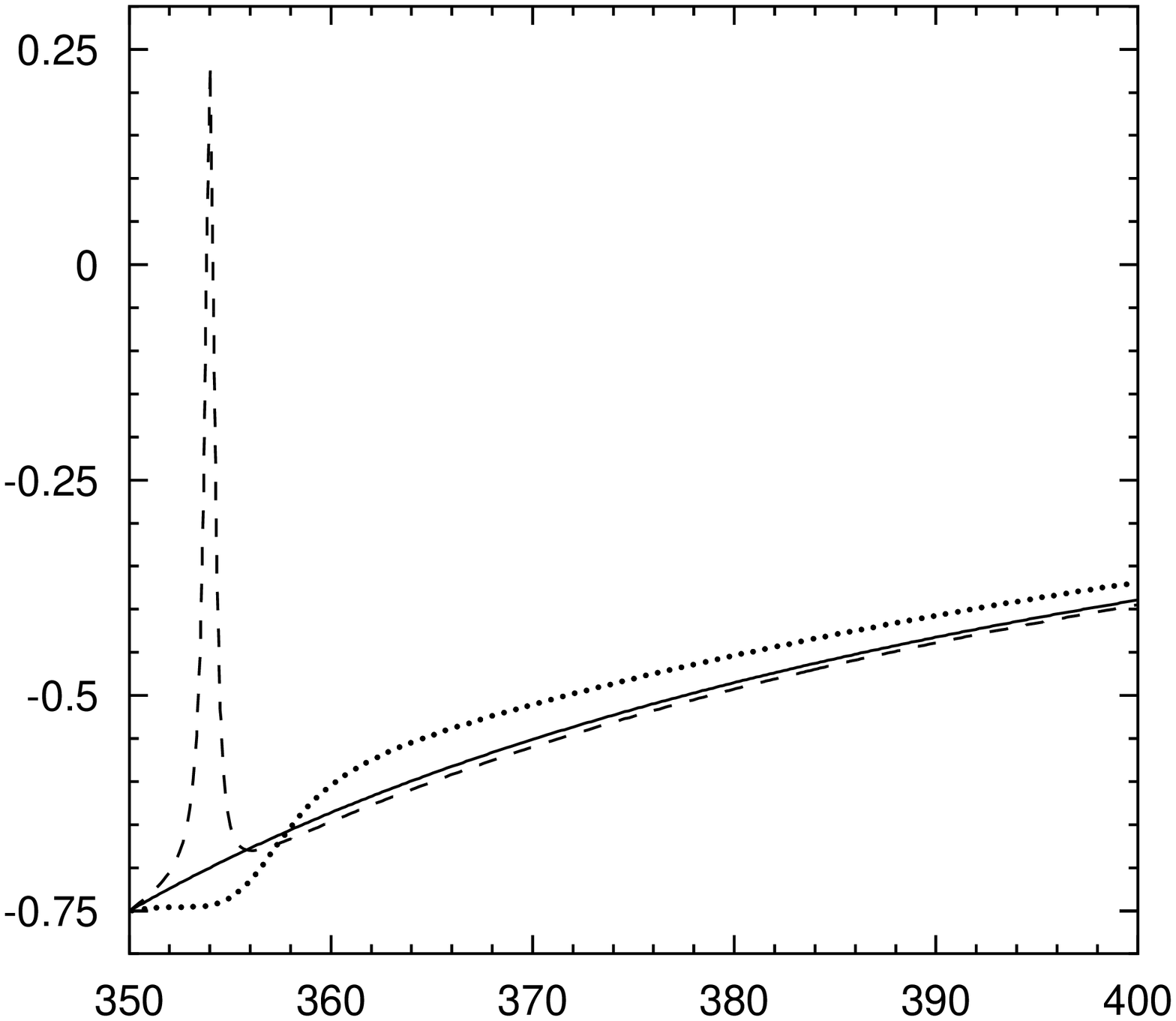} }
\end{picture}
\end{figure}
\par\noindent
\begin{minipage}[t]{1.5cm} {\bf Fig.~1:} \end{minipage}
\begin{minipage}[t]{13.3cm} Expectation value 
$\langle \,{\bf s}_{t}\cdot{\bf s}_{\bar t}\, \rangle$
of the product of $t$ and $\bar t$ spins
in the process $gg\to t\bar t$, plotted as function
of the parton c.m. energy  $\sqrt{\hat s}$.
The solid line is the result
for the background, while the dashed (dotted) line shows
the effect of a scalar (pseudoscalar) Higgs boson with mass
$m_\varphi=354$ GeV and reduced Yukawa couplings
$a=1$, $\tilde a=0$ ($a=0$, $\tilde a=1$).
\end{minipage}\hfill\par
\newpage
\centerline{\bf Figure 2}
\begin{figure}[h]
\unitlength1.0cm
\begin{picture}(15.,9.)
\put( 1.5,0.5){\makebox(4.,1.){$M_{t\bar t}\;\;$[GeV]}}
\put(10.0,0.5){\makebox(4.,1.){$M_{t\bar t}\;\;$[GeV]}}
\put( 1.5,-8.5){\makebox(4.,1.){$M_{t\bar t}\;\;$[GeV]}}
\put( 3.7, 7.1){\makebox(4.,1.){\bf a)}}
\put(12.2, 7.1){\makebox(4.,1.){\bf b)}}
\put( 3.7,-1.9){\makebox(4.,1.){\bf c)}}
\put(-0.9,1.1){
\epsfysize=8.5cm
\epsffile{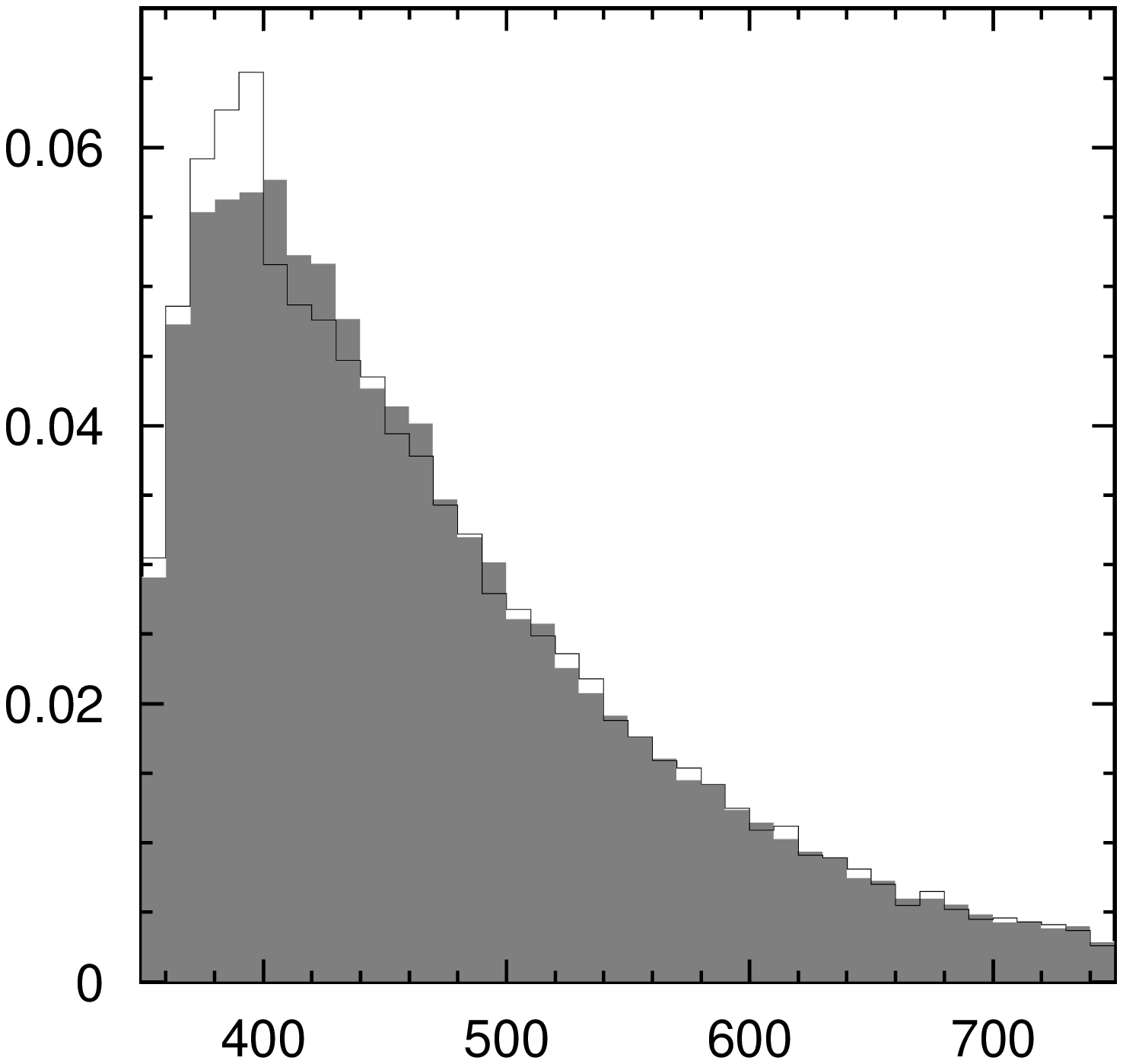} }
\put(7.6,1.1){
\epsfysize=8.5cm
\epsffile{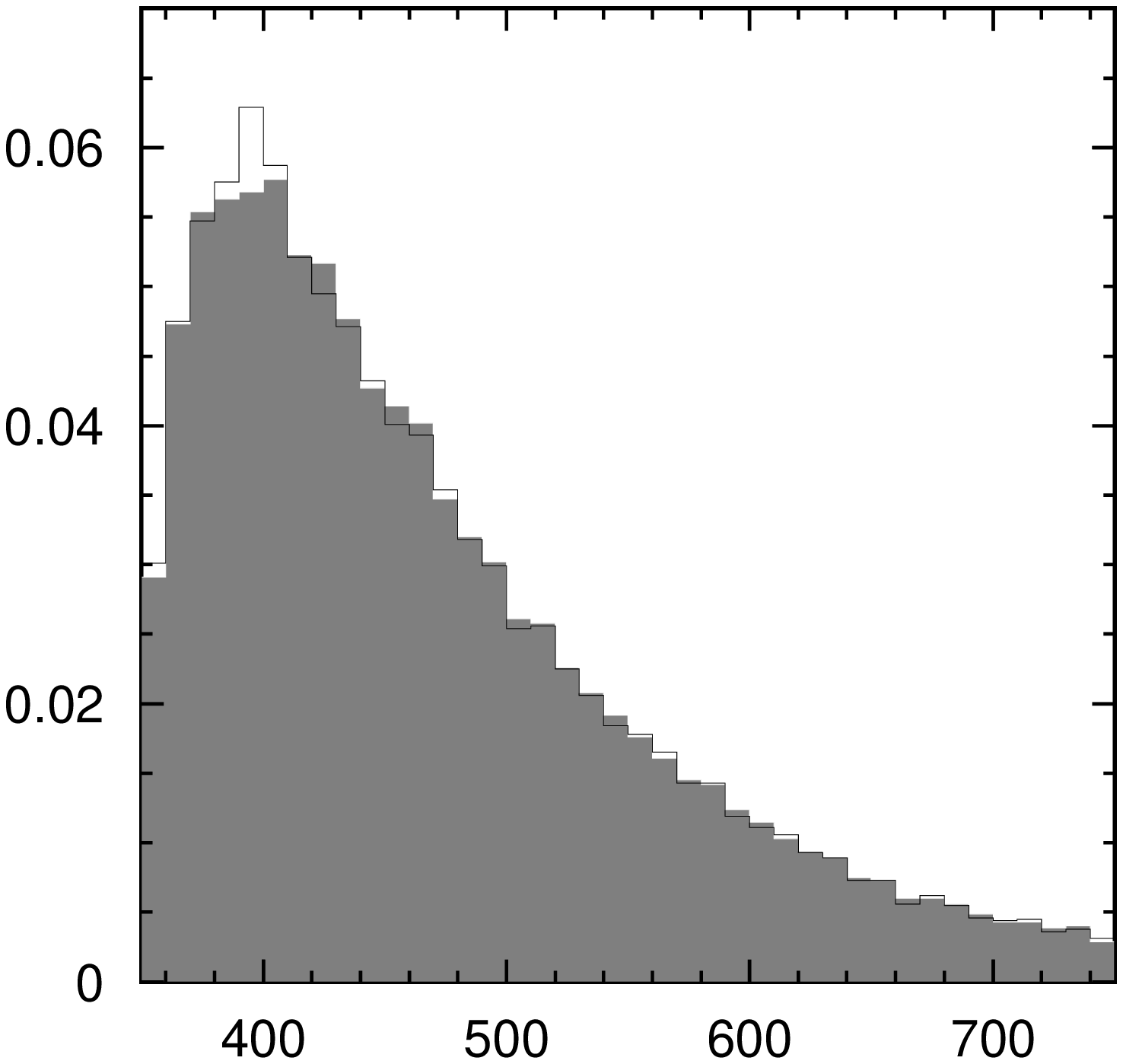} }
\put(-0.9,-7.9){
\epsfysize=8.5cm
\epsffile{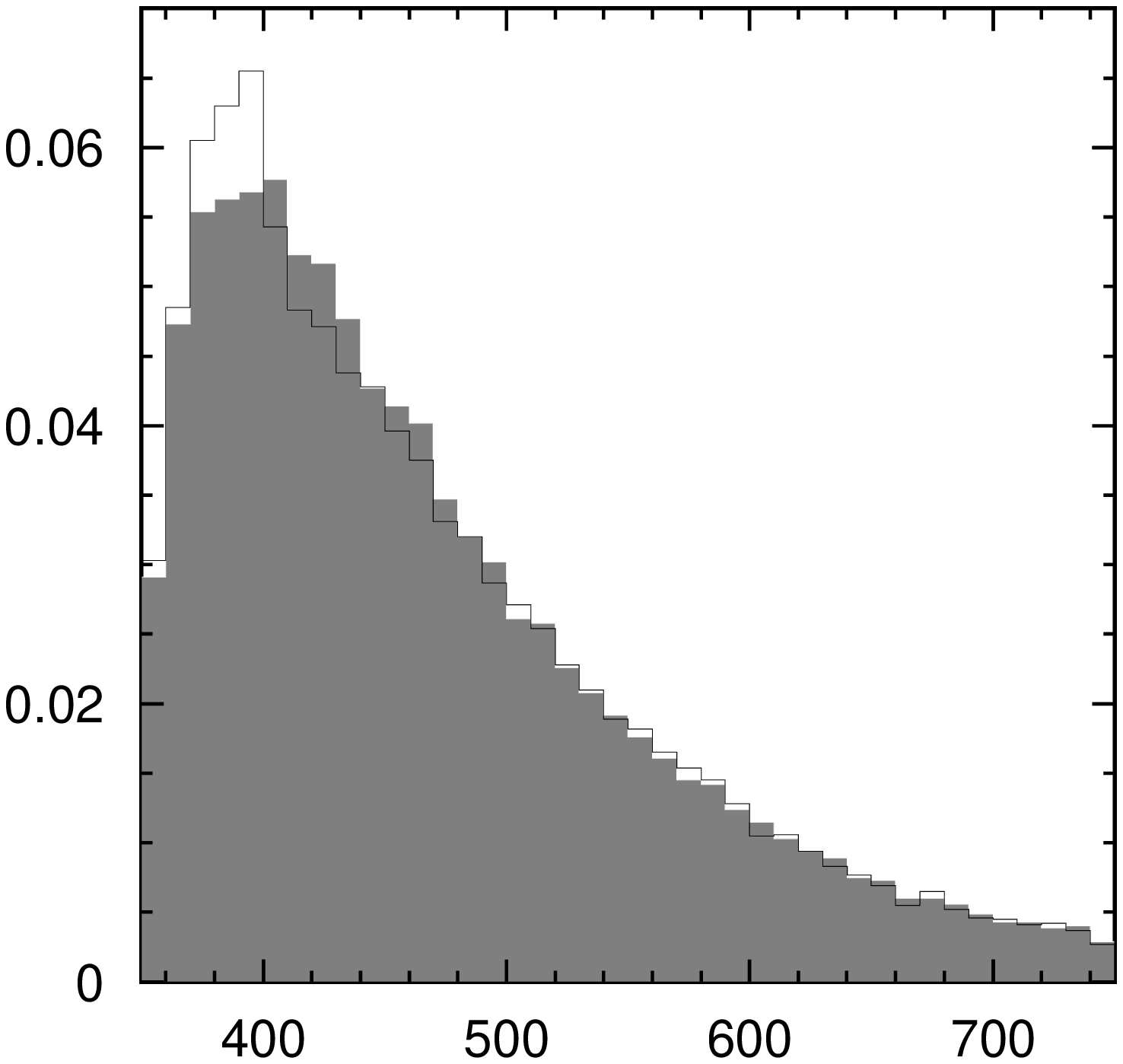} }
\end{picture}
\end{figure}
\par\noindent
\begin{minipage}[t]{7.3cm} \hphantom{text}\hfill \end{minipage}
\hfill \begin{minipage}[t]{7.3cm} {\bf Fig. 2:} 
Normalized $t\bar t$ invariant mass spectrum
for $p\bar p$ collisions at 4 TeV. Shown is the background
contribution (grey) and the effects of a Higgs boson with
mass $m_\varphi\!=\!400$ GeV: {\bf a)} $a\!=\!1$, $\tilde{a}\!=\!0$
(scalar), {\bf b)} $a\!=\!0$, $\tilde{a}\!=\!1$ (pseudoscalar),
{\bf c)} $a=\tilde{a}\!=\!1$ (mixed).
\end{minipage}\par
\newpage
\centerline{\bf Figure 3}
\begin{figure}[h]
\unitlength1.0cm
\begin{picture}(15.,9.)
\put( 1.5,0.5){\makebox(4.,1.){$M_{t\bar t}\;\;$[GeV]}}
\put(10.0,0.5){\makebox(4.,1.){$M_{t\bar t}\;\;$[GeV]}}
\put( 1.5,-8.5){\makebox(4.,1.){$M_{t\bar t}\;\;$[GeV]}}
\put( 3.7, 7.1){\makebox(4.,1.){\bf a)}}
\put(12.2, 7.1){\makebox(4.,1.){\bf b)}}
\put( 3.7,-1.9){\makebox(4.,1.){\bf c)}}
\put(-0.9,1.1){
\epsfysize=8.5cm
\epsffile{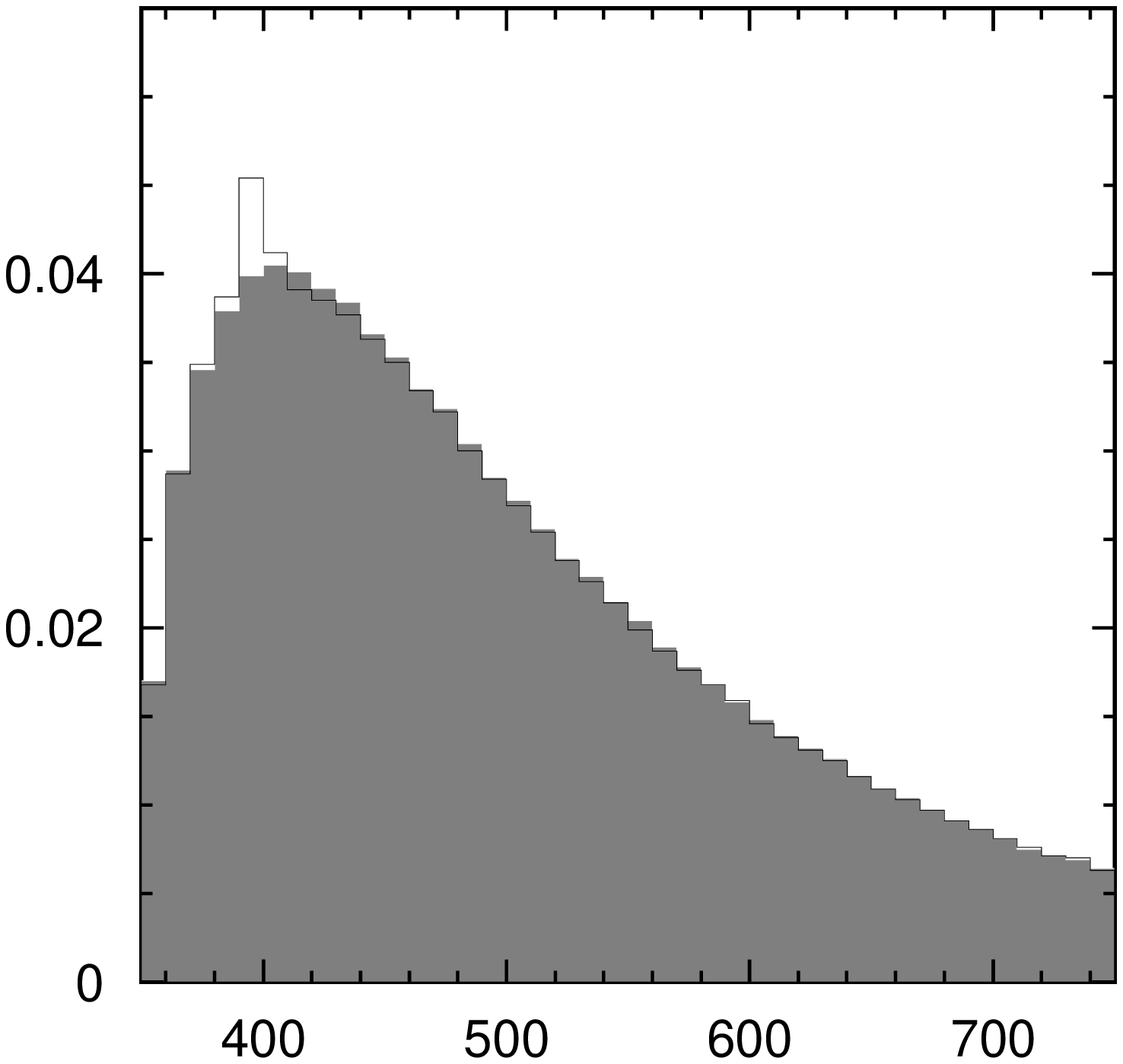} }
\put(7.6,1.1){
\epsfysize=8.5cm
\epsffile{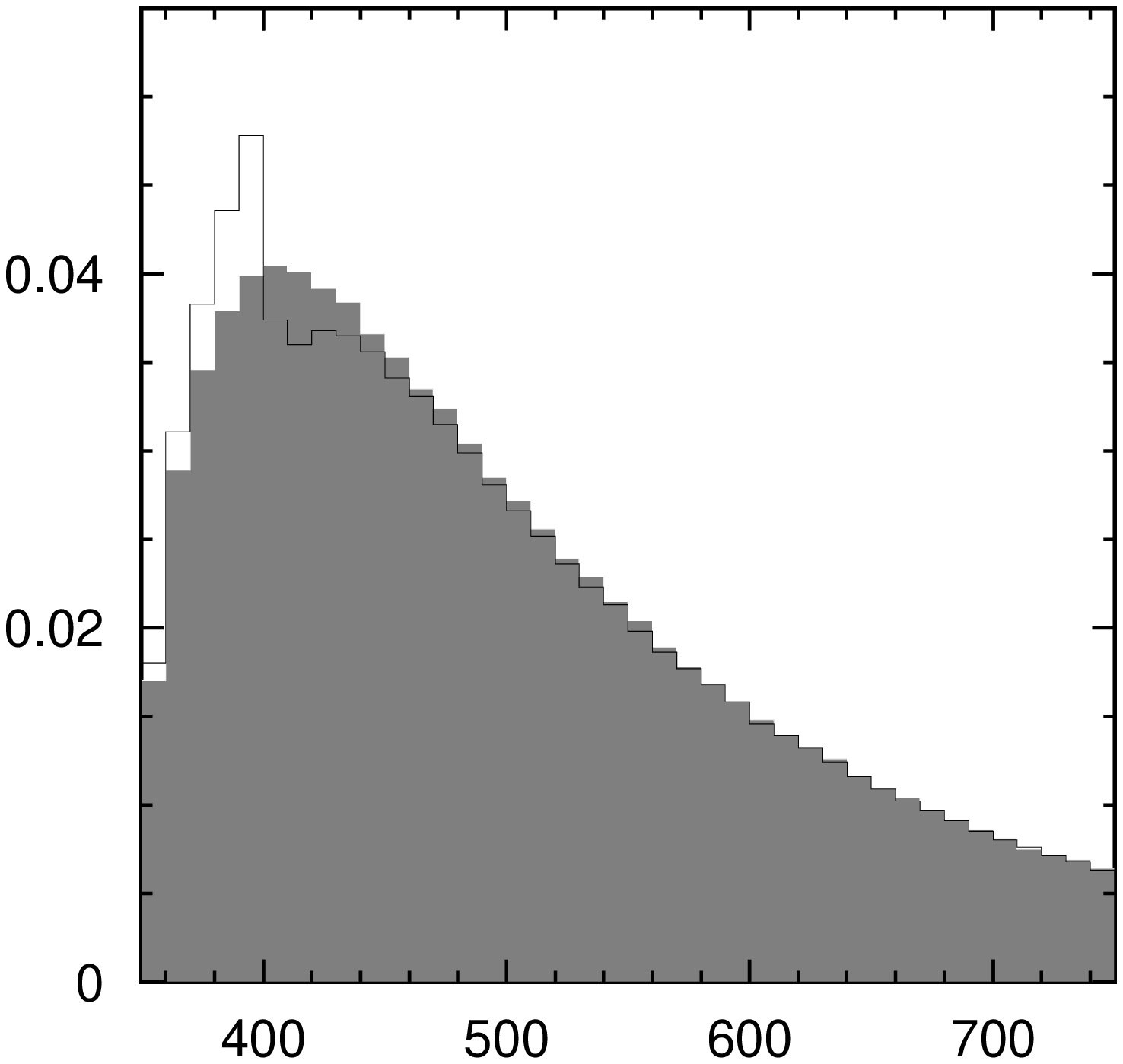} }
\put(-0.9,-7.9){
\epsfysize=8.5cm
\epsffile{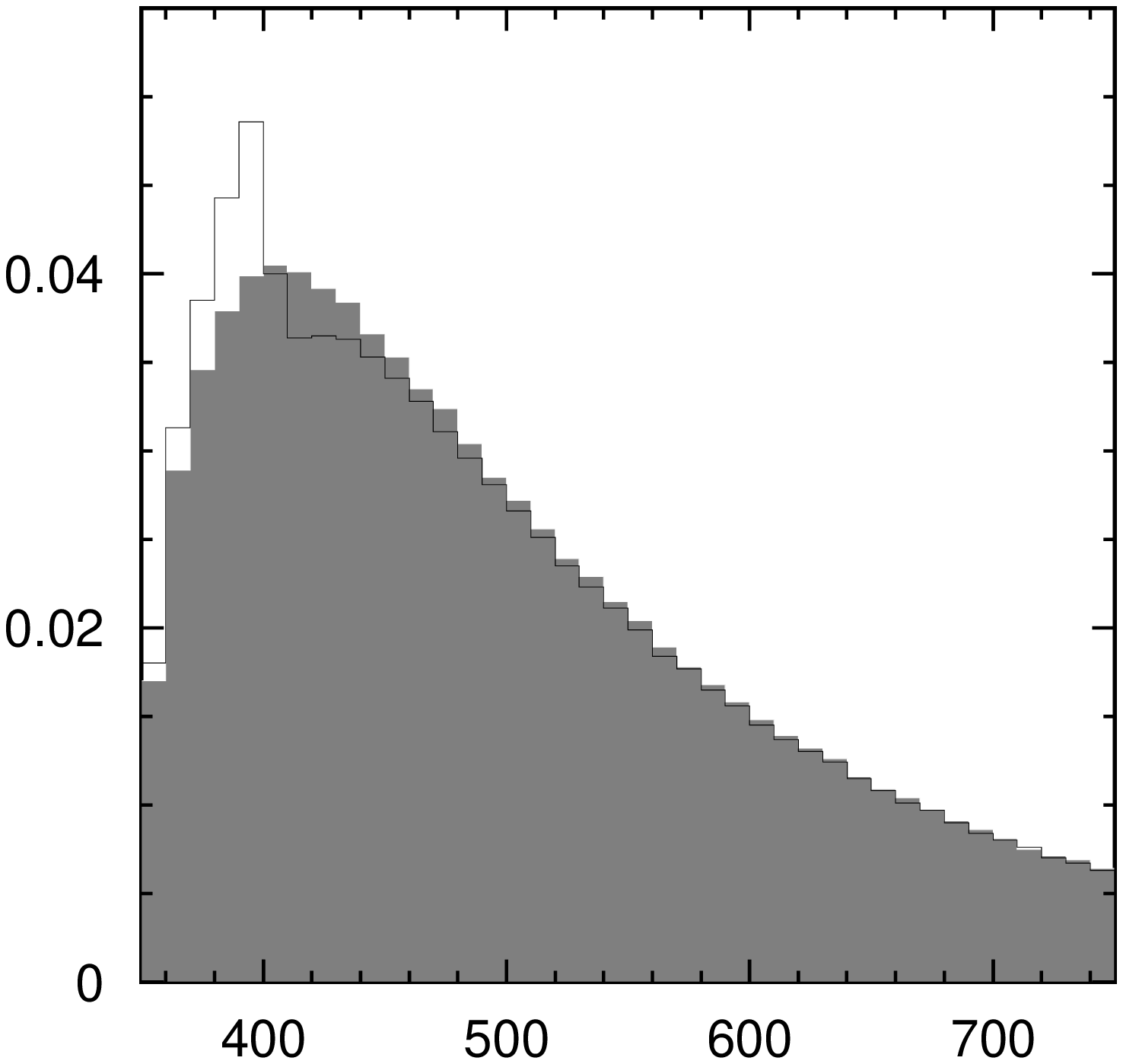} }
\end{picture}
\end{figure}
\par\noindent
\begin{minipage}[t]{7.3cm} \hphantom{text}\hfill \end{minipage}
\hfill \begin{minipage}[t]{7.3cm} {\bf Fig. 3:} 
Same as in Fig.~2, but for $pp$ collisions at 14 TeV. 
\end{minipage}\par
\newpage
\centerline{\bf Figure 4}
\begin{figure}[h]
\unitlength1.0cm
\begin{picture}(15.,17.)
\put( 1.5,8.7){\makebox(4.,1.){$M_{t\bar t}\;\;$[GeV]}}
\put(10.0,8.7){\makebox(4.,1.){$M_{t\bar t}\;\;$[GeV]}}
\put( 1.5,0.2){\makebox(4.,1.){$M_{t\bar t}\;\;$[GeV]}}
\put(10.0,0.2){\makebox(4.,1.){$M_{t\bar t}\;\;$[GeV]}}
\put( 3.7, 15.1){\makebox(4.,1.){\bf a)}}
\put(12.2, 15.1){\makebox(4.,1.){\bf b)}}
\put( 3.7,6.6){\makebox(4.,1.){\bf c)}}
\put(12.2,6.6){\makebox(4.,1.){\bf d)}}
\put(-0.9,9.1){
\epsfysize=8.5cm
\epsffile{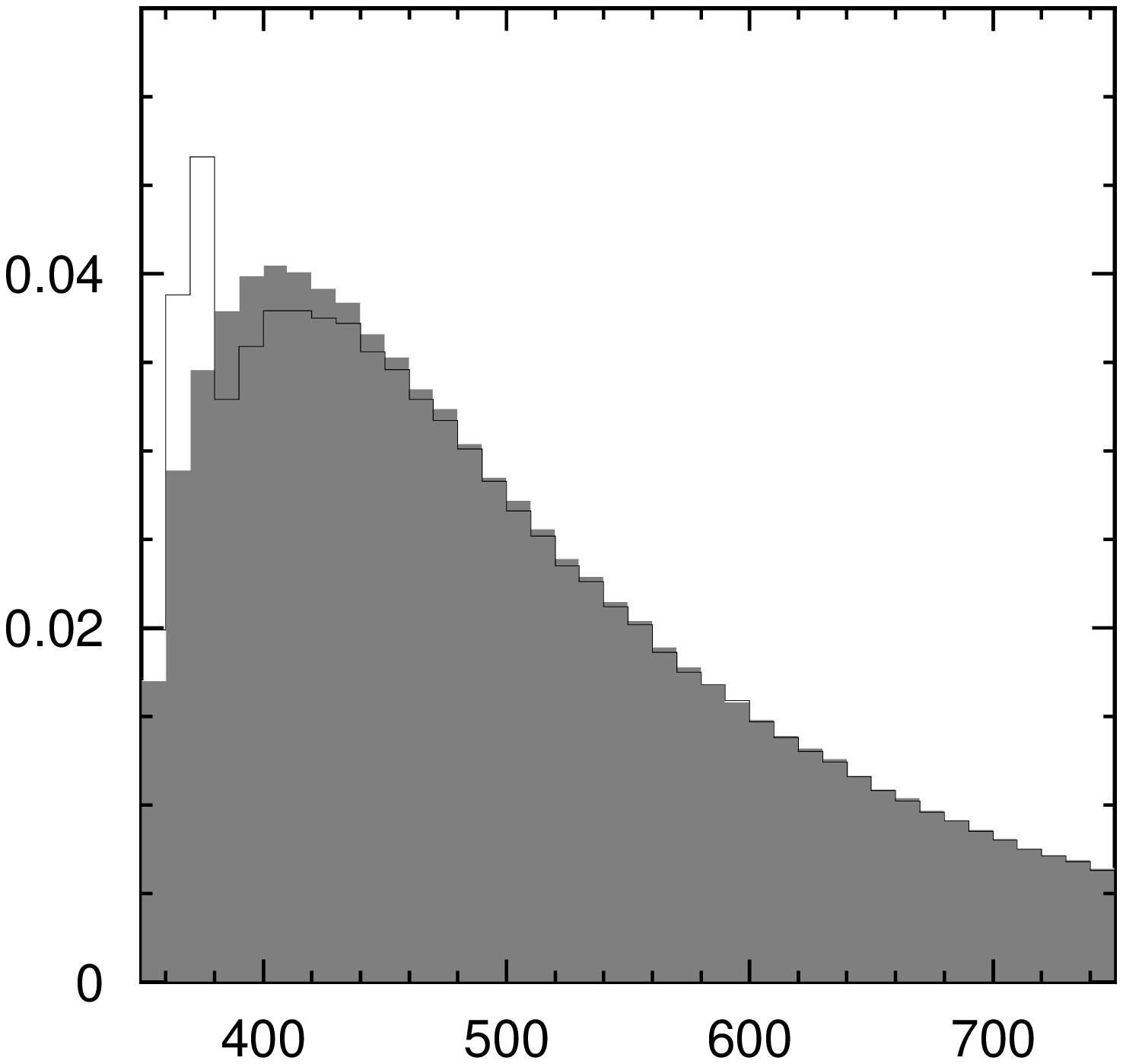} }
\put(7.6,9.1){
\epsfysize=8.5cm
\epsffile{fig3b.eps} }
\put(-0.9,0.6){
\epsfysize=8.5cm
\epsffile{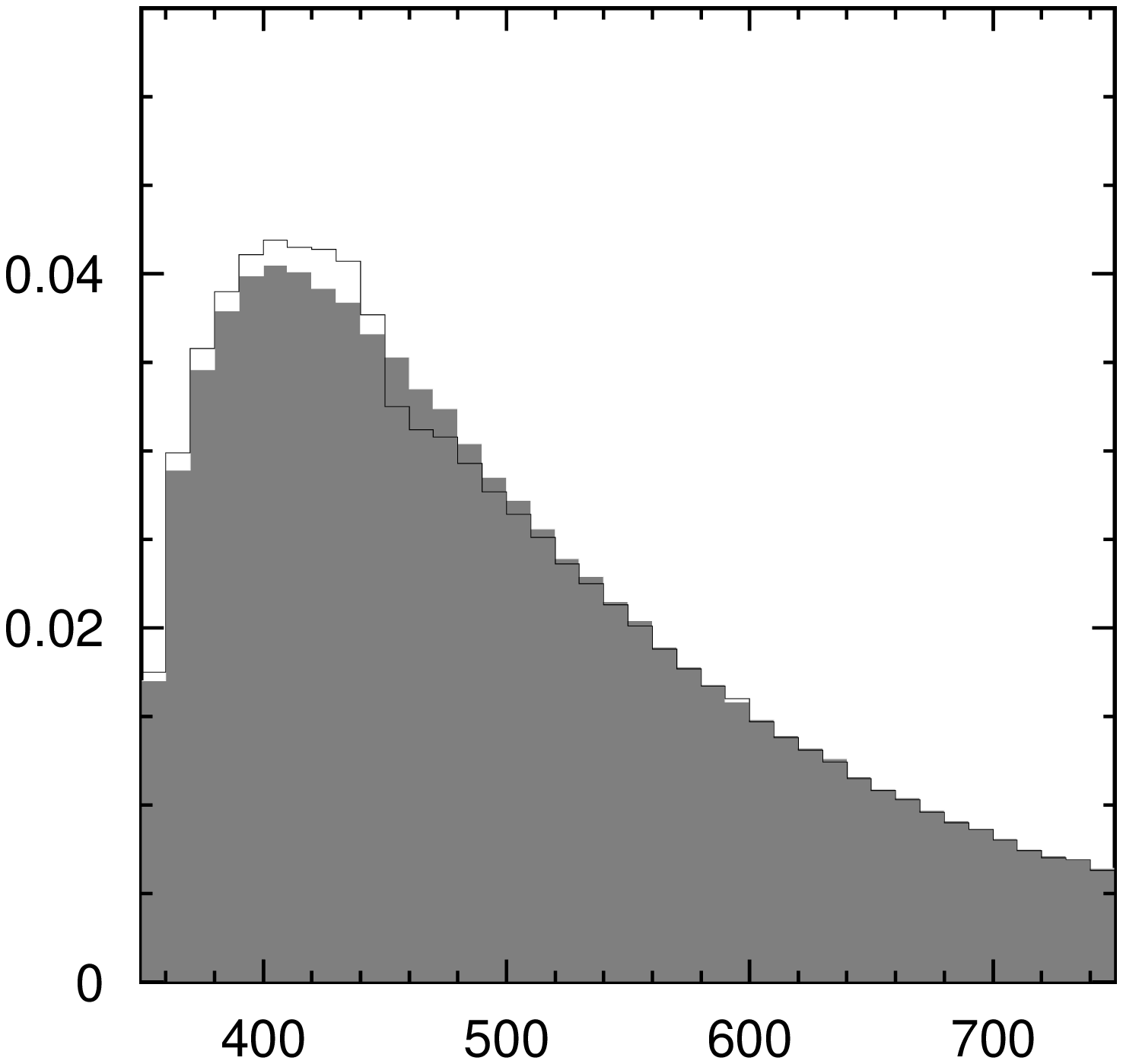} }
\put(7.6,0.6){
\epsfysize=8.5cm
\epsffile{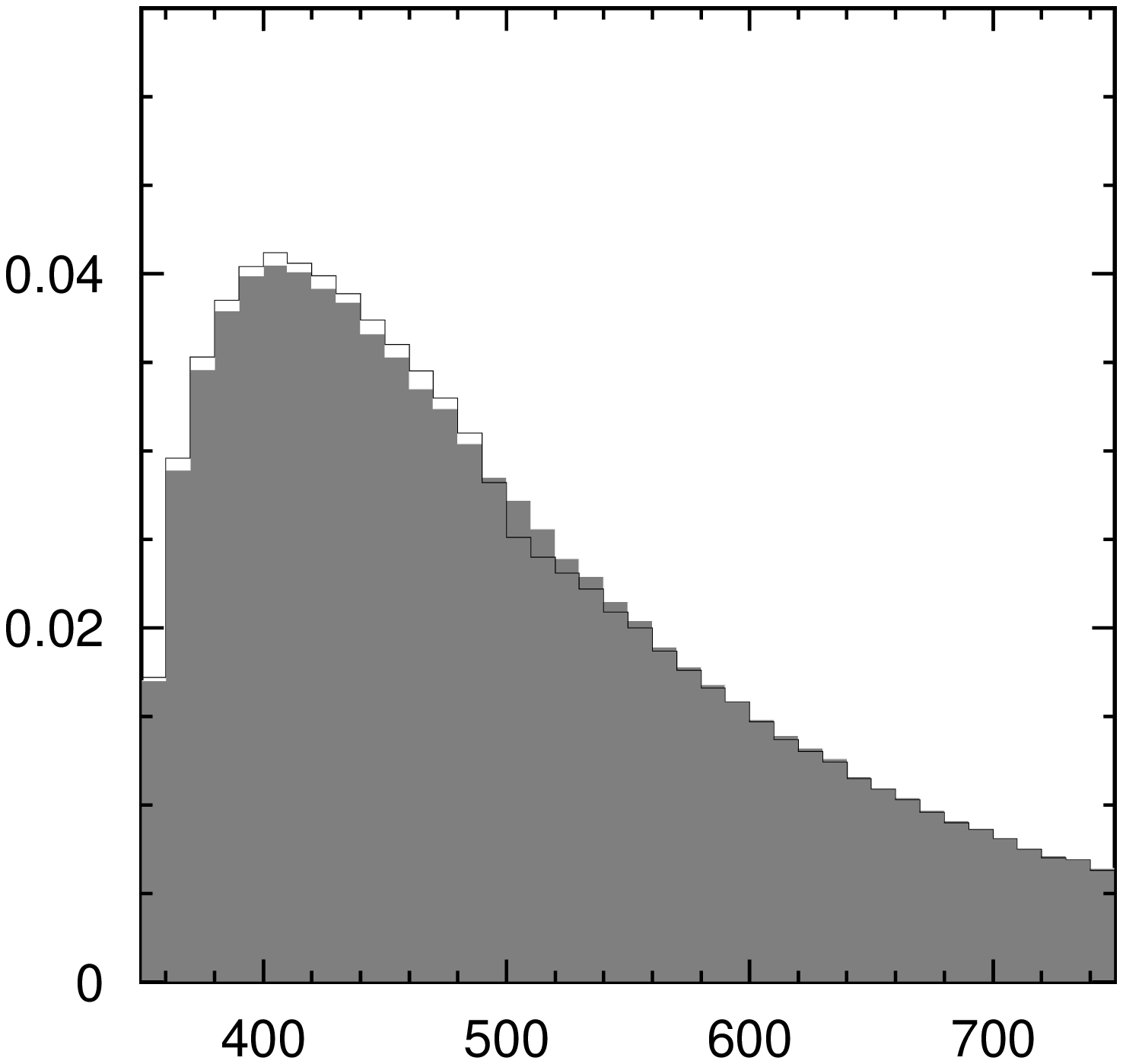} }
\end{picture}
\end{figure}
\par\noindent
\begin{minipage}[t]{1.5cm} {\bf Fig.~4:} \end{minipage}
\begin{minipage}[t]{13.3cm} 
Normalized $t\bar t$ invariant mass spectrum
for $pp$ collisions at 14 TeV. Shown is the background
contribution (grey) and the effects of a pseudoscalar 
Higgs boson with couplings $a=0$, $\tilde{a}=1$.
The different plots {\bf a)}, {\bf b)}, {\bf c)}, {\bf d)}
are for mass values $m_\varphi=375$, 400, 450, 500 GeV,
respectively.
\end{minipage}\hfill\par
\newpage
\centerline{\bf Figure 5}
\begin{figure}[h]
\unitlength1.0cm
\begin{picture}(15.,16.)
\put(-.5,13.3){\makebox(2.,1.){\Large 
$\frac{1}{\sigma}\frac{d\sigma}{d{\cal O}}$}}
\put(7.7,0.6){\makebox(2.,1.){\large $\cal O$ }}
\put(0.2,0.6){
\epsfysize=16.5cm
\epsffile{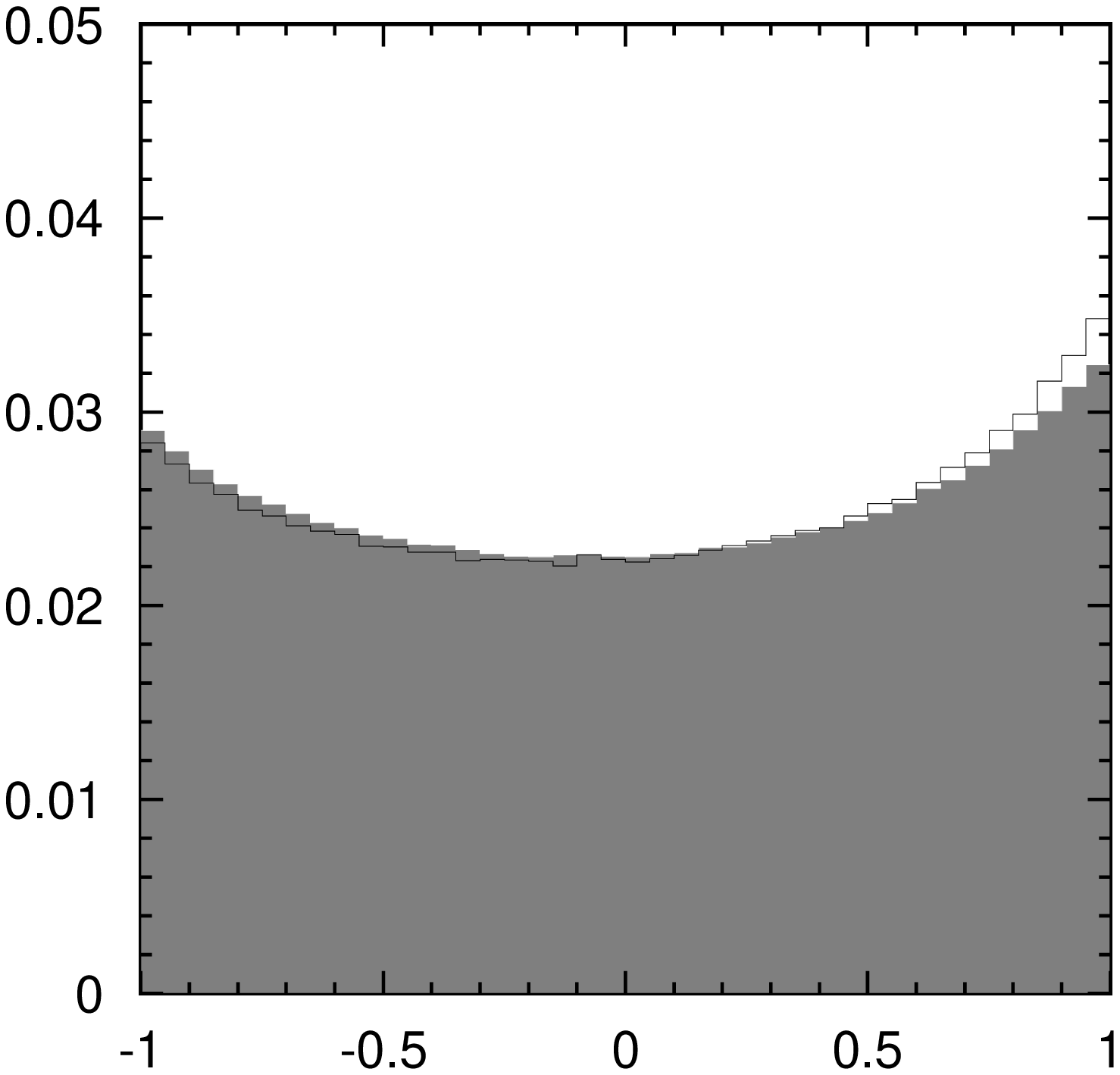} }
\end{picture}
\end{figure}
\par\noindent
\begin{minipage}[t]{1.5cm} {\bf Fig.~5:} \end{minipage}
\begin{minipage}[t]{13.3cm} 
Normalized distribution 
$\sigma^{-1}d\sigma/d{\cal O}$, where
$\cal O$ is the cosine of the angle between the two leptons
in the double lepton channel of $pp$ collisions at
$\sqrt{s}=6$ TeV. The grey area is the contribution
from the background only. The solid line shows the effect of 
a pseudoscalar Higgs boson with mass
$m_\varphi=400$ GeV and couplings $a=0$, $\tilde a=2$.
\end{minipage}\hfill\par
\newpage
\centerline{\bf Figure 6}
\begin{figure}[h]
\unitlength1.0cm
\begin{picture}(15.,16.)
\put(-.5,13.9){\makebox(2.,1.){\large 
$\langle {\cal O} \rangle$}}
\put(7.2,0.6){\makebox(2.,1.){\large $\sqrt{s}\;\;$ [TeV]}}
\put(-0.3,0.6){
\epsfysize=16.5cm
\epsffile{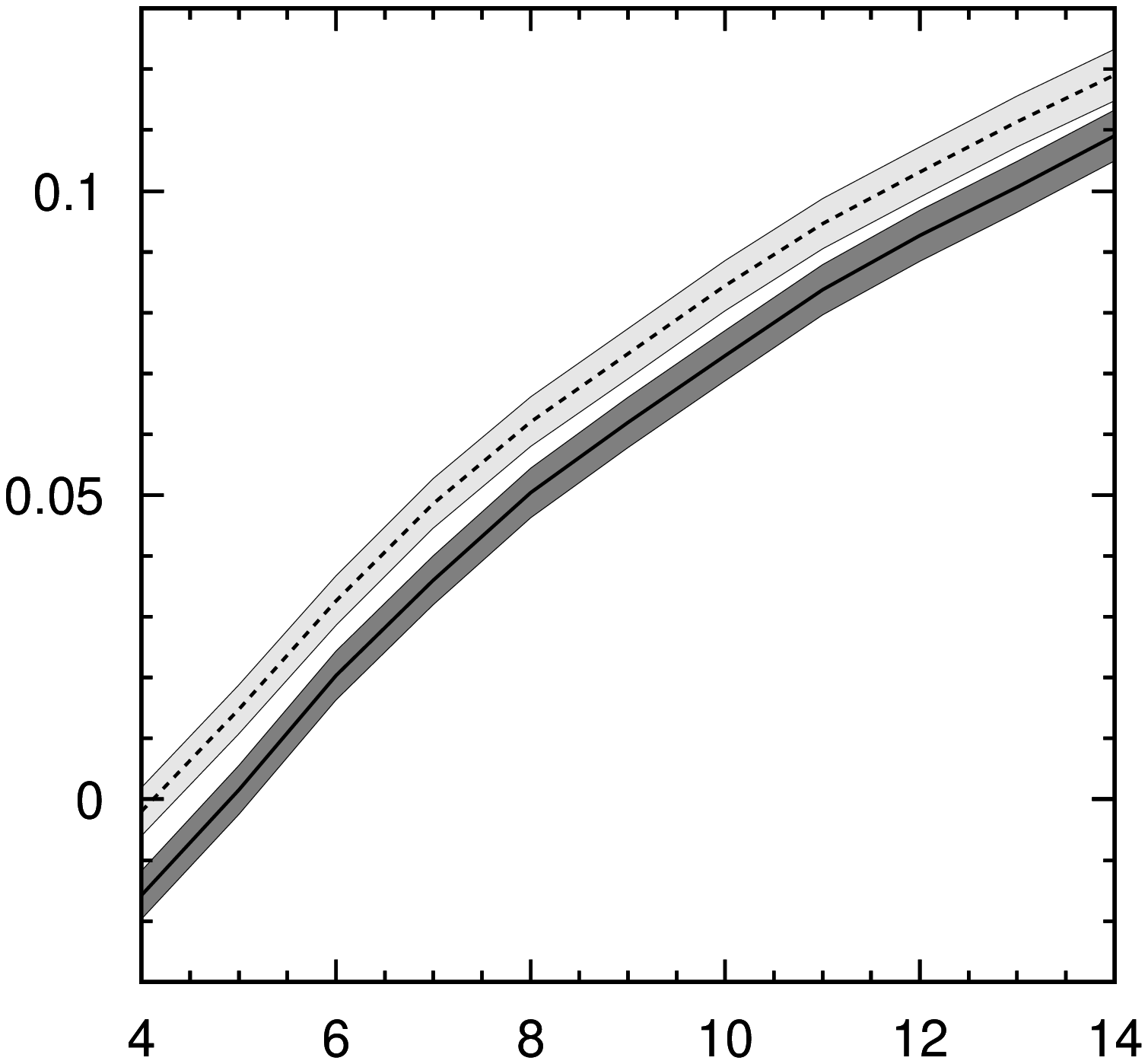} }
\end{picture}
\end{figure}
\par\noindent
\begin{minipage}[t]{1.5cm} {\bf Fig.~6:} \end{minipage}
\begin{minipage}[t]{13.3cm} Expectation value 
$\langle{\cal O} \rangle$ of the cosine of the angle spanned 
by the two leptons in the double lepton channel of $pp$ 
collisions, plotted as function of the c.m.\ energy $\sqrt{s}$.
The solid line and the dark grey band represent the expectation
from the background only, for $2\!\times\!10^5$ events at the 
3 $\sigma$ level. The dashed line with the light grey band was 
obtained with a pseudoscalar ($a=0$, $\tilde a=2$) Higgs boson 
with mass $m_\varphi=400$ GeV.
\end{minipage}\hfill\par

\baselineskip=12pt
\begin{thebibliography}{99}
\bibitem{rev}
For reviews, see for instance
J.~Gunion, H.~Haber, G. Kane and S. Dawson, The Higgs Hunter's Guide
(Addison-Wesley, New York, 1990); \\ Z. Kunszt, hep-ph/9704263 (1997).
\bibitem{Dicus}
D. Dicus, A. Stange, and S. Willenbrock,
Phys. Lett. B {\bf 333}, 126 (1994).
\bibitem{BBra}
W. Bernreuther and A. Brandenburg, Phys. Lett.  B {\bf 314}, 104 (1993);
Phys. Rev. D {\bf 49}, 4481 (1994).
\bibitem{Arens}
T. Arens and L. M. Sehgal,
Phys. Lett. B {\bf 302}, 501 (1993).
\bibitem{Brand}
A. Brandenburg, Phys. Lett. B {\bf 388}, 626 (1996).
\bibitem{Parke}
G. Mahlon and S. Parke, Phys. Rev. D {\bf 53}, 4886 (1996).
\bibitem{Chang}
D. Chang, S. Lee, and A. Soumarokov, Phys. Rev. Lett. {\bf 77}, 1218 (1996).
\bibitem{Will}
T. Stelzer and S. Willenbrock, Phys. Lett. B {\bf 374}, 169 (1996).
\bibitem{corr}
See, for instance, W. Bernreuther, A. Brandenburg, and M. Flesch,
Phys. Rev. D {\bf 56}, 90 (1997), and references cited therein.
\bibitem{Wein}
S. Weinberg, Phys. Rev. D {\bf 42}, 860 (1990).
\bibitem{Georgi}
H. M. Georgi, S. I. Glashow, M. E. Machacek, and D. V. Nanopoulos, \\
Phys. Rev. Lett. {\bf 40}, 692 (1978).
\bibitem{QCD}
In particular the QCD corrections to the lowest order production rates
were computed in [13--15]
and it was found that they increase
the lowest order rates significantly (see also \cite{kumost}).
\bibitem{SH}
A. Djouadi, M. Spira, and P. Zerwas, Phys. Lett. B {\bf 264}, 440 
(1991); \\ D. Graudenz,  M. Spira, and P. Zerwas, Phys. Rev. Lett. 
{\bf 70}, 1372 (1993); \\ S. Dawson, Nucl. Phys. {\bf B359}, 283 (1991).
\bibitem{SH2}
M. Spira, A. Djouadi, D. Graudenz,  and P. Zerwas,
 Nucl. Phys.  {\bf B453}, 17 (1995).
\bibitem{SSH}
M. Spira, A. Djouadi, D. Graudenz,  and P. Zerwas,
Phys. Lett. B {\bf 318}, 347 (1993);
S. Dawson, A. Djouadi, and M. Spira, Phys. Rev. Lett. {\bf 77}, 16 (1996).
\bibitem{kumost}
Z.~Kunszt, S.~Moretti, and W. J. Stirling, Z. Phys. C {\bf 74},
479 (1997).
\bibitem{Gaemers}
K. Gaemers and F. Hoogeveen, Phys. Lett. B {\bf 146}, 347 (1984).
\bibitem{been}
W. Beenakker {\it et al.}, Nucl. Phys. {\bf B411}, 343 (1994).
\bibitem{EW}
The remaining nonresonant Yukawa corrections are of no concern to us 
here; they have been shown to be small in general. See ref.\
\cite{Zhou} for CP-conserving and \cite{Peskin,BBra} for CP-violating 
Higgs boson exchange in two-Higgs doublet models.
The one-loop electroweak corrections  to $g g \to t\bar t$ within the 
SM were computed in \cite{been} and the resulting correction to the 
$p p\to t{\bar t}X$ cross section was found
to be of the order of a few percent.
\bibitem{Zhou}
H. Y. Zhou, C. S. Li, and Y. P. Kuang, Phys. Rev. D {\bf 55}, 4412 (1997).
\bibitem{Peskin}
C.R. Schmidt, M.E. Peskin, Phys. Rev. Lett. {\bf 69}, 410 (1992).
\bibitem{Hara}
Y. Hara, Prog. Theor. Phys. {\bf 86}, 779 (1991).
\bibitem{BNOS}
W. Bernreuther, O. Nachtmann, P. Overmann, and T. Schr\"oder,
Nucl. Phys. {\bf B388}, 53 (1992); {\bf B406}, 516 (E) (1993).
\bibitem{MB}
J. P. Ma and A. Brandenburg,
Z. Phys. C {\bf 56}, 97 (1992).
\bibitem{Haberl}
P. Haberl, O. Nachtmann, and A. Wilch,  Phys. Rev. D {\bf 53},  4875 (1996).
\bibitem{Lad}
G. A. Ladinsky, Phys. Rev. D {\bf 46}, 3789 (1992), {\bf 47}, 3086(E) (1993).
\bibitem{GRV}
M. Gl\"uck, E. Reya, and A. Vogt, Z. Phys. C {\bf 67}, 433 (1995).
\bibitem{MRSCTEQ}
A. D. Martin, W. J. Stirling, and R. G. Roberts, Phys. Lett.
B {\bf 354}, 155 (1995); H. L. Lai {\it et al.} (CTEQ coll.), 
Phys. Rev. D {\bf 51}, 4763 (1995).
\bibitem{BM}
 A. Brandenburg and J. P. Ma, Phys. Lett. B {\bf 298},  211 (1993).
\bibitem{Grz} 
B. Grzadkowski, B. Lampe, and K. J. Abraham, hep-ph/9706489 (1997).
\end{thebibliography}
\end{document}